# Adjoint "quarks" and the physics of confinement

Howard D. Trottier[a][*]

[a]Department of Physics, Simon Fraser University, Burnaby, B.C., Canada V5A 1S6

The quark-antiquark potential and the chromoelectric fields generated by "quarks" in the adjoint representation of four-dimensional SU(2) color are analyzed in the scaling region of the theory. New results with interesting implications for our understanding of the confinement mechanism are presented. In particular, the formation of color-electric flux-tubes between adjoint quarks is demonstrated. The flux-tubes for fundamental and adjoint representation quarks are shown to have very similar cross-sections. This result could imply that the QCD vacuum is dual to a type I superconductor.

Linear confinement in QCD has been well established by lattice simulations of the quark-antiquark ($Q\overline{Q}$) potential. However the physical mechanism underlying confinement has yet to be established, and this continues to represent a fundamental challenge to our understanding of QCD. Valuable information, such as flux-tube formation, has been obtained from studies of the color fields generated by a $Q\overline{Q}$ pair, as this provides a more detailed probe of the system than is obtained from the potential [1–3].

The quark-antiquark potential and the chromo-electric fields generated by "quarks" in the adjoint representation of SU(2) color are analyzed here in the scaling region of the theory. It is widely expected that adjoint quarks should exhibit dynamics with some important differences from quarks in the fundamental representation, thus providing a unique probe of confinement physics [4,5,2,6]. Recently, Michael has obtained some indication that the potential for adjoint quarks may saturate at very large separations $R$, where the energy in the flux-tube exceeds twice the energy of a gluon-adjoint quark bound state (a color-singlet "gluelump") [5]. Further evidence of adjoint quark screening is obtained here by comparing the distribution of fields generated by quarks $Q_j$ in the adjoint ("isospin" $j = 1$) and fundamental ($j = 1/2$) representations.

The SU(2) lattice theory with standard Wilson action is used. Given a set of links $U_\mu(x)$, a corresponding set of fuzzy links $\tilde{U}^1_\mu(x)$ is constructed according to [7]:

$$\tilde{U}^1_{\mu \neq 4}(x) = \mathcal{N}\Big[ c\, U_\mu(x) + \sum_{\nu \neq \pm\mu, \pm 4} U_\nu(x) U_\mu(x+\hat{\nu}) U^\dagger_\nu(x+\hat{\mu}) \Big], \quad (1)$$

where $c$ is a positive constant, and $\mathcal{N}$ is chosen so that $\det \tilde{U}^1 = 1$. Twenty iterations of this fuzzing procedure, with $c = 2.5$, are used. Wilson loops $\tilde{W}_j$ in the two representations are given by

$$\tilde{W}_j(R,T) \equiv \frac{1}{2j+1} \text{Tr} \left\{ \prod_{l \in L} \mathcal{D}_j[\tilde{U}_l] \right\}, \quad (2)$$

where $\mathcal{D}_j[\tilde{U}_l]$ is an appropriate representation of the link ("fuzzy" spatial links and "ordinary" time-like links). $\tilde{W}_1 = (4\tilde{W}^2_{1/2} - 1)/3$ is used.

The $Q_j \overline{Q}_j$ field strengths are obtained from plaquette correlators. Results are reported here for the component of the color-electric field in the direction $\hat{x}$ parallel to the $Q_j \overline{Q}_j$ axis

$$\mathcal{E}_j(x) \equiv -\frac{\beta}{a^4}\Big[ \langle \tilde{W}_j \tfrac{1}{2}\text{Tr}\, U_{14}(x) \rangle / \langle \tilde{W}_j \rangle - \langle \tfrac{1}{2}\text{Tr}\, U_{14} \rangle \Big], \quad (3)$$

where $U_{14}$ is a plaquette with sides along the $\hat{x}$ and time-like axes. In the continuum limit $\mathcal{E}_j$ reduces to the expectation value of $\frac{1}{2}\sum_a (\vec{E}^a \cdot \hat{x})^2$ in the presence of a $Q_j \overline{Q}_j$ pair, after vacuum subtraction [1].

---

[*]Work supported in part by the Natural Sciences and Engineering Research Council of Canada.

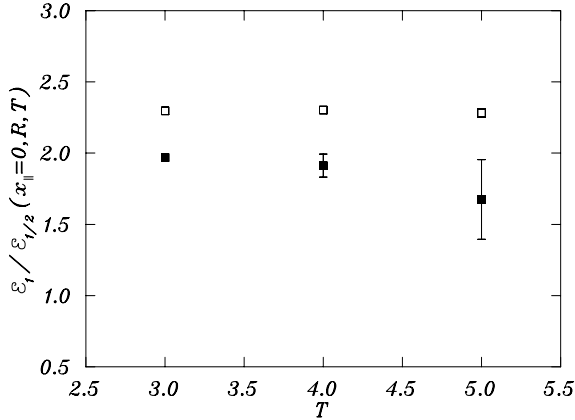

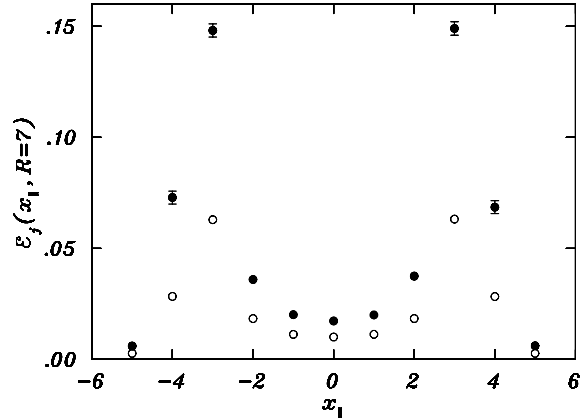

Figure 1. Time-dependence of the ratio of fields $\mathcal{E}_1/\mathcal{E}_{1/2}$ at the center of the Wilson loop for $R = 3$ (open box) and $R = 5$ (filled box).

Figure 2. Fields versus $x_\parallel$ at $R = 7$ for $j = 1$ (•) and $j = 1/2$ (○).

Analytical integrations are made on the timelike links in the Wilson loop [8], and over the four links in a plaquette [9]. This provides for a reduction in errors in $\mathcal{E}_j$ by a factor of as much five. It is also advantageous to use $\langle U \rangle \simeq \langle W_j U(x_R) \rangle / \langle W_j \rangle$ [1], which is found to be well satisfied with $x_R = \sqrt{3} \times 8$ lattice units from the center of the Wilson loop.

The calculations were done on a $16^4$ lattice at $\beta = 2.4$. More than 10,000 sweeps were used for thermalization. 1,800 measurements were made of fuzzy Wilson loops of sizes $R \times T$ from $3 \times 1$ to $8 \times 5$. 6,000 further measurements were then made of loops with $R = 7$ and 8. 100 heat bath sweeps were made between measurements, yielding integrated autocorrelation times $\tau_{\text{int}} \lesssim 0.5$. Estimates of the statistical errors were obtained using the jackknife method. The "time-dependent" estimate of the potential $V_{\text{est},j} = \ln[\tilde{W}_j(R, T-1)/\tilde{W}_j(R, T)]$ becomes independent of $T$ for $T \gtrsim 3$, within statistical errors. The same is true of the field strength, as shown in Figure 1, although the statistical errors in the fields are much larger than in the potential (useful results for $R = 7$ and 8 could only be obtained for $T = 3$).

The effect of screening of adjoint quarks is apparent in the potential data. At $R \times T = 8 \times 4$, for example, $V_{\text{est},1}/V_{\text{est},1/2} = 2.47 \pm 0.02$. Michael has obtained results for larger $R$ [5]; at $R = 12$, $V_1/V_{1/2}$ is between about 2.1 and 2.3 (the uncertainty comes from an estimate of the systematic errors in the extrapolation to $T \to \infty$). These results are to be compared with the ratio of Casimirs $j(j+1)$ of the two representations, equal to $8/3 = 2.66$.

Screening is more evident in the field strength. Figures 2 and 3 show the fields as functions of position $x_\parallel$ in the plane of the Wilson loop, using data at $T = 3$. The quarks are located at $x_\parallel = \pm R/2$. Figure 4 shows the ratio of fields at the center of the Wilson loop as functions of $R$ (at $T = 3$). Near the quarks, the ratio of field strengths $\mathcal{E}_{j=1}/\mathcal{E}_{j=1/2}$ is close to the Casimir ratio $8/3$, for all $R$. In the region between the two quarks the ratio of fields falls below $8/3$, reaching a minimum at $x_\parallel = 0$. The ratio is significantly reduced at large $R$. At $R = 5$, for example, $\mathcal{E}_{j=1}/\mathcal{E}_{j=1/2} = 1.97 \pm 0.02$ at $T = 3$, and $1.91 \pm 0.08$ at $T = 4$. Quantitatively similar results for the ratio of fields at a given $R$ are found at all points in the plane perpendicular to the $Q_j \overline{Q}_j$ axis (see Figs. 5 and 6, described below). This suggests that there is a reduction in the color-electric flux in the region between the adjoint quarks.

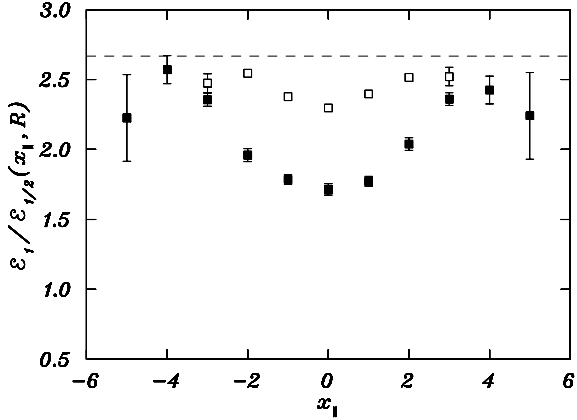

Figure 3. Ratio of fields versus $x_\parallel$ for $R = 3$ (open box) and $R = 7$ (filled box). The dashed line shows the ratio of Casimirs.

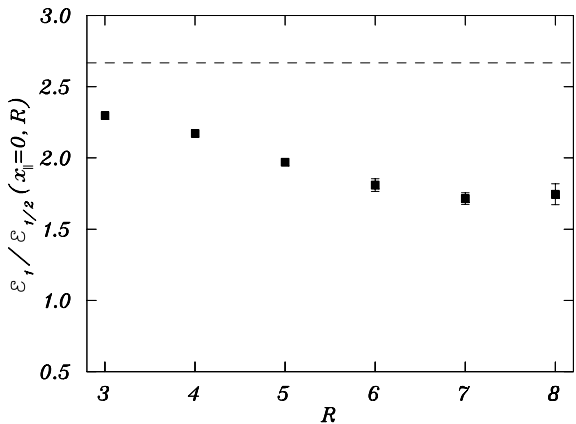

Figure 4. Ratio of fields in the center of the Wilson loop versus $R$.

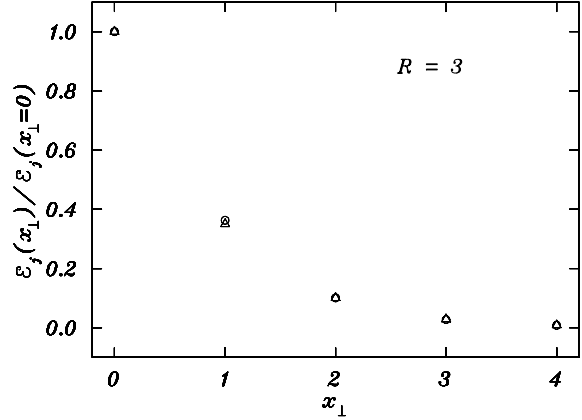

Figure 5. Fields $\mathcal{E}_j$ versus $x_\perp$ at $R = 3$ for $j = 1$ ($\triangle$) and $j = 1/2$ ($\circ$). The fields have been normalized to their values at $x_\perp = 0$.

One naively expects that adjoint quarks become completely screened at sufficiently large $R$, where it becomes energetically favorable for the flux-tube to fission, leading to the formation of a pair of quark-gluon bound states ("gluelumps") [5]. One would therefore expect the fields in between the quarks to approach vacuum values at large $R$. The fact that the ratio of fields $\mathcal{E}_{j=1}/\mathcal{E}_{j=1/2}$ in between the quarks falls well below the ratio of Casimirs as $R$ increases can be interpreted as evidence of vacuum screening. However, even at the largest $R$ probed here, $\mathcal{E}_{j=1}$ is greater than $\mathcal{E}_{j=1/2}$. Nevertheless, the data presented here suggests that the local adjoint field strength exhibits screening which, in some regions, is more pronounced than is indicated by the potential. This is presumably due to the fact that the potential averages the fields over all space. A calculation on larger lattices of the plaquette correlators at larger $T$ and $R$ is needed in order to determine whether the adjoint quark fields become completely screened at large separations. Unfortunately, the methods used here appear to be inadequate for this task.

Results for $\mathcal{E}_j$ as a function of position $x_\perp$ perpendicular to the $Q_j \overline{Q}_j$ axis are shown in Figs. 5

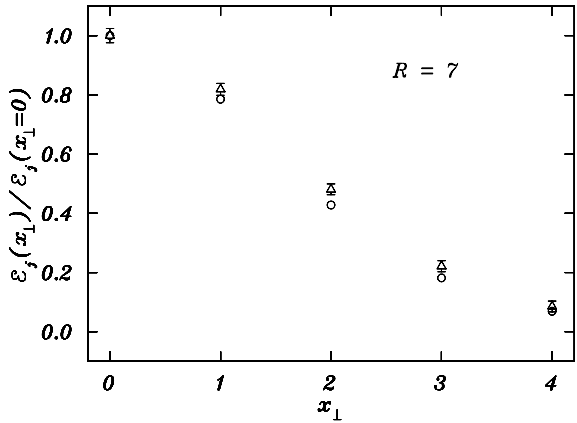

Figure 6. Fields $\mathcal{E}_j$ versus $x_\perp$ at $R = 7$ for $j = 1$ ($\triangle$) and $j = 1/2$ ($\circ$).

and 6. The fields exhibit a "penetration depth" of a few lattice spacings. The fundamental and adjoint quark flux-tubes appear to have very similar cross-sections, despite appreciable screening of the fields in between the adjoint quarks at large $R$.

This result has implications for the dual superconductor model of confinement [10]. In type II superconductors a domain wall formed between normal and superconducting regions has negative surface energy, so that many normal regions are created, each carrying an elementary quantum of flux. In type I superconductors however a domain wall has positive surface energy, hence as few normal regions as possible are created [11]. If the dual superconductor picture is correct, then the fact that adjoint quarks form a single flux-tube, with a cross-sectional structure similar to the flux-tube formed by fundamental quarks, may suggest that the QCD vacuum is dual to a type I superconductor.

These results also suggest a connection between the confinement mechanism in QCD in both three and four dimensions, and in three-dimensional QED. The fields generated by $Q\overline{Q}$ pairs in several representations of three-dimensional SU(2) and U(1) lattice theories were calculated in Ref. [2]. The fields were found to be restricted to a flux-tube whose cross-section is approximately independent of the representation. It is interesting to note that Abelian magnetic monopoles are present in all of these theories.